\documentclass[aps,prb,twocolumn,groupedaddress,showpacs]{revtex4}
\usepackage{epsfig}
\usepackage[dvipsnames,usenames]{color}

\begin{document}

\title{Determinant Quantum Monte Carlo Study of the
Enhancement of d-wave Pairing by Charge Inhomogeneity}

\author{Rubem Mondaini$^1$, Tao Ying$^{2,3}$, Thereza Paiva$^{1}$,  and Richard T.~Scalettar$^2$}

\affiliation{$^1$Instituto de Fisica, Universidade Federal do Rio de
Janeiro Cx.P. 68.528, 21941-972 Rio de Janeiro RJ, Brazil}

\affiliation{$^2$Physics Department, University of California, Davis,
California 95616, USA}

\affiliation{$^3$Department of Physics, Harbin Institute of Technology, Harbin 150001, China}

\begin{abstract}
Striped phases, in which spin, charge, and pairing correlations vary
inhomogeneously in the CuO$_2$ planes, are a known experimental feature
of cuprate superconductors, and are also found in a variety of numerical
treatments of the two dimensional Hubbard Hamiltonian.  In this paper we
use determinant Quantum Monte Carlo to show that if a stripe density
pattern is imposed on the model, the $d$-wave pairing vertex is
significantly enhanced.  We attribute this enhancement to an increase in
antiferromagnetic order which is caused by the appearance of more
nearly half-filled regions when the doped holes are confined to
the stripes.  We also observe a $\pi$-phase shift in the magnetic
order.
\end{abstract}

\maketitle

\section{INTRODUCTION}

Cooper pairs in conventional BCS superconductors are typically envisioned
to have a large spatial extent characterized
by the coherence length $\xi$, which
is many hundreds of lattice spacings in elemental, metallic
superconductors.  At the other extreme is the BEC regime, where much
smaller Cooper pairs form into bosonic particles which subsequently can
Bose-Einstein condense into a superfluid phase.
The crossover between the BCS and BEC limits has been
much explored\cite{randeria12}.

If real space pairing on small length scales is favorable energetically,
a natural question to ask is why more than two fermions do not bind
together.  Indeed, the competition of such ``phase separation" with
superconductivity was a central theme of investigation in early studies of
Hamiltonians like the two dimensional fermion Hubbard
\cite{poilblanc89,schulz90,giamarchi90,fye90,moreo91,becca00,su96,gehlhoff96,zitzler02,aichhorn05,machida89}
and $t$-$J$
\cite{emery90,hellberg01,gimm00,putikka00,shih98,Martins00}
models in the context of cuprate superconductors, which have short
coherence length.  Upon doping away from one particle per site, a
spatial division into half-filled AF regions and areas where the hole
concentration is high occurs.  The tendency to phase separation was
found to be especially great in the $t$-$J$ model, and somewhat less so in
the Hubbard model.  Roughly speaking, for small hole doping in the strong coupling limit such phase separation can be
envisioned to arise because it minimizes the energy by preserving the
largest number of antiferromagnetic bonds. On the other hand, when $t>>J$ and the amount of holes is large, phase separation was shown
to occur as to minimize the doped hole kinetic energy \cite{hellberg01}.

A compromise between complete phase separation into distinct two
dimensional regions, and spatial homogeneity which would be favored by
longer range Coulomb interactions, \cite{emery,low} is the possibility that the doped
particles form quasi-one dimensional patterns in which hole-rich and
hole-poor regions alternate.  Magnetic domain lines were observed in
inhomogeneous Hartree-Fock studies very early in the history of cuprate
superconductivity.  `Charge domain lines' were first reported in a
multi-band model which included both the copper $d$ and oxygen $p$
orbitals \cite{zaanen89}, and subsequently in the single band Hubbard
Hamiltonian \cite{inui91,yang91}.

Such striped patterns have also been observed
experimentally\cite{tranquada95,tranquada98}.  In
La$_{1.6-x}$Nd$_{0.4}$Sr$_x$CuO$_4$ there is a suppression of
superconductivity at $x=1/8$ associated with a tilt pattern of oxygen
tetrahedra and charge/spin domain walls.  Away from $x=1/8$ domain wall
ordering is weaker, with coexisting superconductivity.

The relation between pairing and stripes is still controversial.
The formation of charge stripe-order is related to an increase in  the resistance perpendicular to the CuO planes, frustrating the formation of  bulk superconductivity,
as denoted by a sharp decrease of the critical temperature\cite{li2007}, $T_c$, resulting in values as low as $4$~K.
As stripes are perpendicularly stacked in each successive CuO plane, if the Josephson coupling between stripes in the same plane is negative the coupling between stripes in different planes is destroyed \cite{berg2007,berg2009}.

In LaBaCuO, on
the other hand, angle-resolved spectroscopy and transport measurements
suggest a positive correlation between 2D pairing and stripes
\cite{valla06,tranquada08}.

There is an order of magnitude drop \cite{li2007,tranquada08,hucker11} in the in-plane resistivity, $\rho_{ ab}$, when the spin order occurs at $T_{so} \sim 42$~K. Furthermore, there are indications of true 2D superconductivity for $T<T_{BKT} \approx 16$~K when $\rho_{ ab}$ goes to 0 \cite{li2007,tranquada08}.

 A recent exciting development is the direct evidence of stripes in YBa$_2$Cu$_3$O$_y$.
Nuclear magnetic resonance measurements show that high magnetic fields induce charge order, \cite{tao-wu11} with the same period of four lattice spacings  as in LaBa based cuprates \cite{tranquada04}. Zero-field  diffraction with resonant soft \cite{ghiringhelli12,achkar12} and hard \cite{chang12} x-ray scattering, observed incommensurate charge order with a period around 3.2 lattice spacings.  These experiments seem to indicate that stripes are an intrinsic phenomena in the cuprates, and that at least incipient charge ordering can be seen even in compounds with larger critical temperatures.

This has led to a large set of theoretical
studies of charge ordered patterns to refine and improve upon the
initial mean field treatment
\cite{oles12,arrigoni02,arrigoni03,martin05,aryanpour06,tsai06,loh07,yao07,aryanpour07,mishra08,tsai08,doluweera08}.
Given the complexity of the question of superconductivity even in the
homogeneous model, it is not surprising that the details of the
interplay with stripe formation should be challenging.

A recent calculation within the dynamical cluster approximation
(DCA)\cite{maier10} provided detailed information concerning pairing
correlations amidst static stripes, and revealed a rich competition
between an enhancement of the pairing interaction and a suppression of
the noninteracting susceptibility.  The latter effect occurs as a
consequence of the formation of Mott regions away from the stripes, and
a resulting suppression of quasiparticle weight.  The two effects
combine to give a non-monotonic evolution of the transition temperature
with the stripe modulation strength.

In this paper we undertake determinant Quantum Monte Carlo (DQMC)
studies of stripe formation in the two dimensional Hubbard model.  The
DQMC method complements the DCA approach, working in real space rather
than momentum space.  It is possible to study somewhat larger clusters
with DQMC, with, however, an off-setting greater
restriction to the accessible
temperatures.  The key results of our work are an enhancement of
antiferromagnetic order by charge modulation, and a $\pi$-phase shift
above a critical threshold of the stripe potential.  Accompanying this
larger magnetic order is a stronger signal of $d$-wave pairing in the
associated superconducting vertex, although we are not able to lower the
temperature enough to cross below the transition temperature.

\section{Hamiltonian and Methodology}

We study a two dimensional repulsive Hubbard Hamiltonian
in which stripes are introduced externally
via a raised site energy $V_0$ on a set of rows of period ${\cal P}$,
\begin{eqnarray}
\label{Hamiltonian}
{\hat\mathcal H}&=&-t\sum_{\langle {\bf i\,j} \rangle \, \sigma}
(c_{{\bf i}\sigma}^\dagger c_{{\bf j}\sigma}^{\phantom{\dagger}} +
c_{{\bf j}\sigma}^\dagger c_{{\bf i}\sigma}^{\phantom{\dagger}} )
+ U \sum_{\bf i} n_{{\bf i}\uparrow} n_{{\bf i} \downarrow}
\nonumber
\\
&-&\mu \sum_{{\bf i}}
(n_{{\bf i}\uparrow} + n_{{\bf i}\downarrow})
+V_0 \sum_{i_y \in {\cal P}}
(n_{{\bf i}\uparrow} + n_{{\bf i}\downarrow})
\label{hamiltonian}
\end{eqnarray}
Here $t=1$ is the intersite fermion hopping between
near neighbor sites ${\bf i},{\bf j}$ on a square lattice, $U$ is
an onsite repulsion, $\mu$ is a global chemical potential,
and $V_0$ is an additional on-site
energy imposed on a set of rows
${\bf i}=(i_x,i_y)$ with
${\rm mod}(i_y,{\cal P})=0$.
When ${\cal P}=4$ this choice produces the spin and charge patterns
postulated based on neutron scattering studies
\cite{tranquada95,tranquada98} and
shown to arise in density matrix renormalization group (DMRG)
studies on the $t$-$J$ Hamiltonian \cite{white98}.

This Hamiltonian does not, of course, address the issue of
{\it spontaneous} stripe formation in a translationally invariant
system, nor does it acknowledge the tendency of charge
domain walls to fluctuate.
Nevertheless it allows us to examine the nature of spin
and pairing correlations in the presence of a set of pre-formed
lines of reduced charge density, and is an appropriate approximate model
in the limit where the energy scale of stripe formation is
considerably greater than that of pairing.

\begin{figure}[t] 
\epsfig{figure=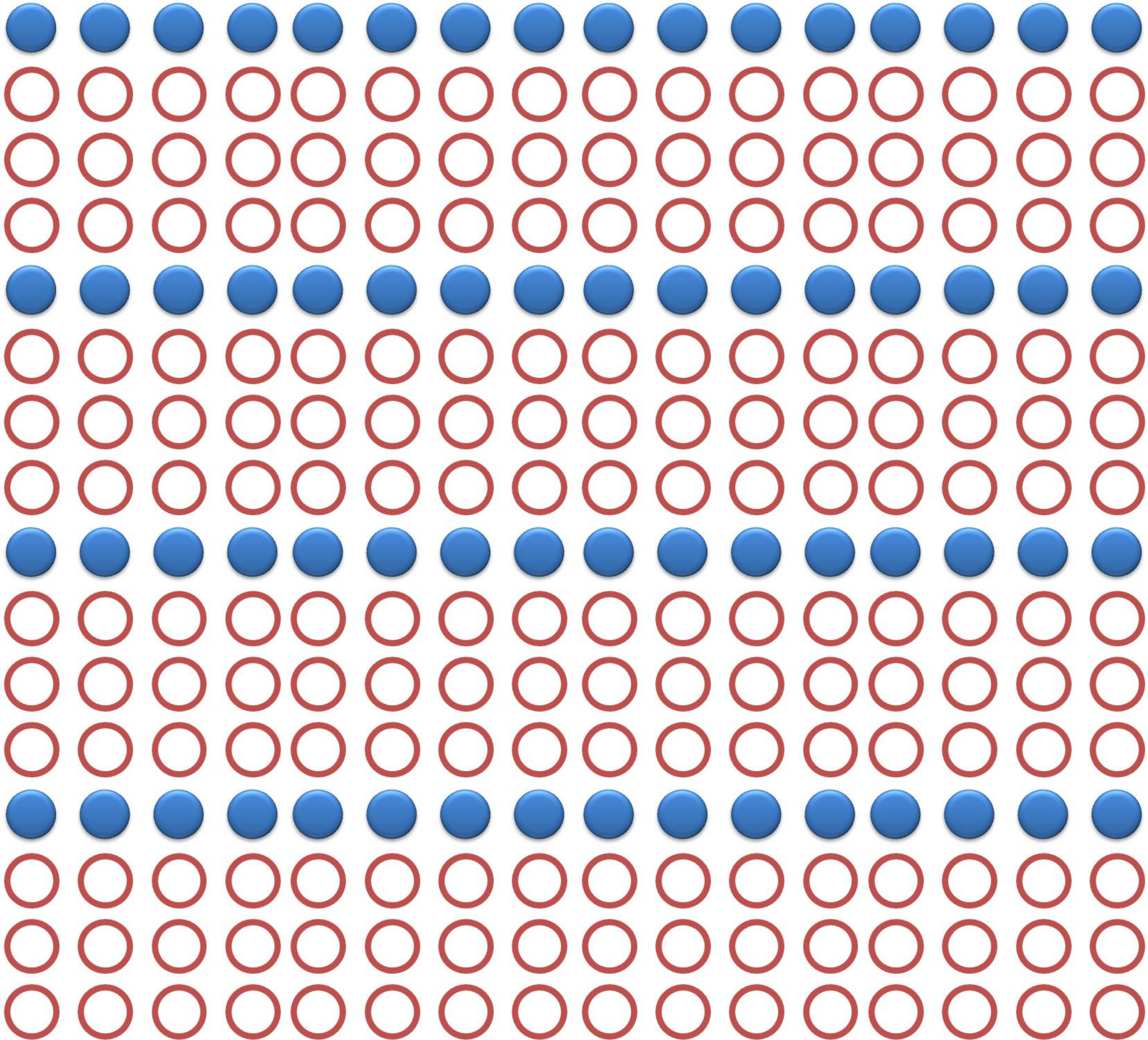,
width=5.0cm,clip}
\caption{(Color online)  16x16 lattice with period ${\cal P}=4$, sites with $V_0$ active are depicted in blue (filled cricles), whereas
the interstripe sites are red (empty circles).
\label{stripesketch}
}
\end{figure}

Most of our results will be for 16x16 lattices with ${\cal P}=4$,
so that each stripe (row with $V_0$ term active, blue filled circles in Fig.~\ref{stripesketch})
is separated by three rows where the $V_0$ term
is not present (empty circles in Fig.~\ref{stripesketch}).
The entire 16x16 site system accomodates four stripes
for this ${\cal P}=4$ case.
We will also analyze
finite size effects and present
a smaller amount of data for pairing correlations at other
periodicities ${\cal P}$.  Most of our results will be for a total
density $\rho=0.774$ averaged over the entire lattice.  This value
was chosen because it allows for the exploration of a broad range
of densities on the stripe and between stripes, and also because
certain experimentally observed characteristics of the striped phase,
such as the `$\pi$ phase shift' of the spin correlations when
traversing a stripe, are absent at some other fillings.

In this ${\cal P}=4$ case, we expect the charge order to be modulated along the  y direction with the same period four, and the spin order has   a period twice as large as a result of the `$\pi$-phase shift' (see below),  also observed in experiments. Therefore,  X-ray and neutron diffraction experiments are expected to show peaks at ${\bf Q}_{CO}=\frac{2\pi}{a} (0;\pm2\delta)$ and ${\bf Q}_{SO}=\frac{2\pi}{a} (0.5;0.5\pm \delta)$ \cite{tranquada95,tranquada96,hucker11}, respectively. The value of $\delta$ is  doping dependent in the La$_{2-x}$Ba$_x$CuO$_4$ family \cite{hucker11} and is assumed to be 1/8 for doping higher than 0.125\cite{yamada98,abbamonte} in agreement with the stripe sketch in Fig.~\ref{stripesketch}.
Some experiments, however, displayed incommensurate charge stripes with $\delta$ smaller than expected \cite{kim,hucker11}.

Our methodology is determinant Quantum Monte Carlo (DQMC)
\cite{bss81,white89a}.  In this approach, the quartic interaction term
is replaced by a coupling of the local $z$ component of spin to an
auxiliary field\cite{foot1}.
The fermion degrees of freedom are integrated out
analytically, leaving a Monte Carlo over the auxiliary field.  In the
process of eliminating the interaction term the inverse temperature
$\beta$ is discretized.  We choose the discretization mesh $\Delta \tau
= 1/8t$.  The resulting Trotter errors are typically only a few percent,
and their elimination is of consequence only if subtle changes in short
range correlation functions are of interest \cite{elena,paiva11}.  In DQMC,
the Trotter errors are typically smaller than accessible statistical
errors on long range spin, charge and pairing correlations at the lowest
temperatures.

\begin{figure}[t] 
\epsfig{figure=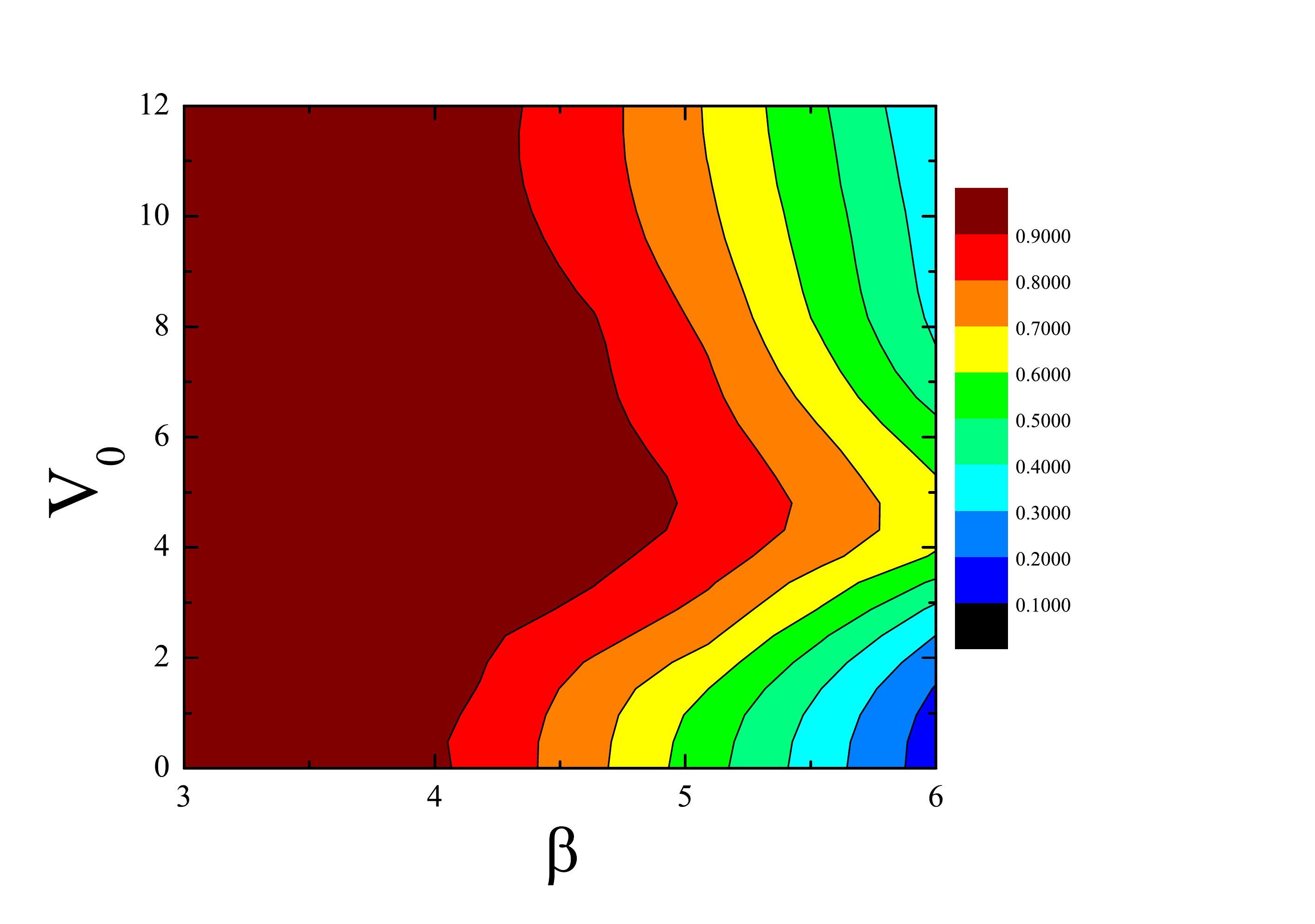,
width=10.0cm,clip}
\vspace{-0.5cm}
\caption{(Color online) Contour plot of the sign as a function of $V_0$ and $\beta$ for a 16x16 lattice with $U=4$ and $\rho=0.774$. The contours are basically vertical, indicating that the sign problem is independent of $V_0$. However for $V_0 \sim 4$ there is a modest improvement in the sign. This is the potential for which the interstripe density goes through half-filling.
\label{sign}
}
\end{figure}

The stripe potential $V_0$ breaks particle-hole symmetry so that there
is a sign problem\cite{loh90} for all fillings.  Thus we cannot obtain
ground state properties as can be accessed, for example, in the
half-filled homogeneous system or the attractive Hubbard Hamiltonian
at any filling. A contour plot
of the sign for $\rho=0.774$ and $U=4$ for a 16x16 lattice,
is shown in Fig. \ref{sign}.  It can be seen that
the sign problem is no worse than in the doped, homogeneous
case where one can, already discern significant trends concerning the
superconducting correlations \cite{white89b}.

We will show results for the spin and
(d-wave) pair correlations,
\begin{eqnarray}
c_{\rm spin}({\bf i})&=&
\langle S^-_{{\bf j}+{\bf i}} S^+_{\bf j} \rangle
\hskip0.6in
S^+_{\bf j} = c^\dagger_{{\bf j}\uparrow}
c^{\phantom{\dagger}}_{{\bf j}\downarrow}
\nonumber \\
c_{d \,{\rm pair}}({\bf i})&=&
\langle \Delta^{\phantom{\dagger}}_{d\, {\bf j}+{\bf i}}
\Delta^{\dagger}_{d\,{\bf j}} \rangle
\nonumber \\
\Delta^{\dagger}_{d\,{\bf j}} &=&
c^\dagger_{{\bf j}\uparrow} (c^{\dagger}_{{\bf j}+\hat x\downarrow}
-c^{\dagger}_{{\bf j}+\hat y\downarrow}
+c^{\dagger}_{{\bf j}-\hat x\downarrow}
-c^{\dagger}_{{\bf j}-\hat y\downarrow} )
\label{equaltime}
\end{eqnarray}
The quantities $c_{\rm spin}({\bf i})$ and
$c_{d \, {\rm pair}}({\bf i})$
do not have complete translation invariance owing to
the presence of the stripes.  They will depend on the row index
$i_y$,  for example whether mod$(i_y,{\cal P})=0$ so the row
has a reduced density, or, for
mod$(i_y,{\cal P}) \neq 0$ on the distance from the
reduced density stripe.

In a phase with long range spin or pairing order, the appropriate
correlation function would approach a constant value asymptotically
as $|{\bf i}| \rightarrow \infty$.  Indeed, precisely this is seen in the
ground state of the half-filled Hubbard  model which has long range
anti-ferromagnetic order\cite{hirsch85,white89a},
and the attractive Hubbard model at a range of
fillings\cite{scalettar89}.
It is still an open issue whether $c_{d \,{\rm pair}}({\bf i})$ is nonzero
at long distances when the homogeneous Hubbard model is doped away from
half-filling, because the sign problem
prevents attaining the ground state in the doped case.

Given this limitation on the simulations, it is important
to develop methods which
extract the maximal useful information about the tendency
to order at temperatures above the putative superconducting phase
transition.  To this end, one introduces a somewhat
more sensitive measure of pairing by considering
the pair-field susceptibility $P_d$ and its associated vertex.
To define $P_d$ we first extend the definition of the
equal time pair
correlation function $c_{d \,{\rm pair}}({\bf i})$
to allow the insertion and removal
of  the Cooper pair to be separated in imaginary time.
$P_d$ is the sum over all spatial sites ${\bf i}$ and
integral over all imaginary time $\tau$
separations of $c_{d \, {\rm pair}}({\bf i},\tau)$.
\begin{eqnarray}
c_{d \, \rm pair}({\bf i},\tau)&=&
\langle \Delta^{\phantom{\dagger}}_{d\,{\bf j}+{\bf i}}(\tau)
\Delta^\dagger_{d\,{\bf j}}(0) \rangle
\nonumber \\
\Delta^\dagger_{d\, {\bf j}}(\tau) &=&
e^{\tau H} \Delta^\dagger_{d\, {\bf j}}(0) e^{-\tau H}
\nonumber \\
P_d &=&  \sum_{\bf i} \int_0^\beta  c_{d\,{\rm pair}}({\bf i},\tau)
\,\, d\tau
\label{pairsusc}
\end{eqnarray}

We also define the uncorrelated pair field susceptibility
$\overline{P}_d$ which instead computes the expectation values of
pairs of operators {\it prior} to taking the product,
with expressions like
$\langle \,\, c_{{\bf i}+{\bf j}\, \downarrow}^{\phantom{\dagger}}(\tau)
\,c_{{\bf i}+{\bf j}\,\uparrow}^{\phantom{\dagger}}(\tau)
\,c_{{\bf j}\, \uparrow}^{\dagger}(0)
\,c_{{\bf j}\,\downarrow}^{\dagger}(0) \,\, \rangle$ which
appear in evaluating the $P_d$ in Eq.~\ref{pairsusc}
being replaced by
$\langle \,\, c_{{\bf i}+{\bf j}\, \downarrow}^{\phantom{\dagger}}(\tau)
\,c_{{\bf j}\,\downarrow}^{\dagger}(0) \,\, \rangle
\,\,\langle \,\, c_{{\bf i}+{\bf j}\, \uparrow}^{\phantom{\dagger}}(\tau)
\,c_{{\bf j}\,\uparrow}^{\dagger}(0) \,\, \rangle$.

$P_d$ includes both the renormalization of the propagation of the
individual fermions as well as the interaction vertex between them,
whereas $\overline{P}_d$ includes only the former effect.
In short,
in DQMC, the averaging over the Hubbard-Stratonovich field
replaces the interaction with the one body potential by the original
electron-electron interactions, so that the order of averaging and
multiplying the operators can be used to control which many body effects
are included.

By evaluating both $P_d$ and $\overline{P}_d$ we are able to extract
\cite{white89b} the interaction vertex $\Gamma_d$,
\begin{eqnarray}
\Gamma_d = {1 \over P_d}
- {1 \over \overline{P}_d} \,\,.
\label{vertex}
\end{eqnarray}
If $\Gamma_d \overline P_d < 0$, the associated pairing
interaction is attractive.  In fact, rewriting Eq. \ref{vertex}  as,
\begin{eqnarray}
P_d = \frac{\overline P_d}{1+\Gamma_d \overline P_d}
\label{stonerlike}
\end{eqnarray}
suggests that $\Gamma_d \overline P_d \rightarrow -1$ signals a
superconducting instability.  This is the analog of the familiar
Stoner criterion $U \chi_0 = 1$ which arises
from the random phase approximation expression $\chi=\chi_0/(1-U\chi_0)$
for the interacting magnetic susceptibility $\chi$
in terms of the noninteracting $\chi_0$.
We will discuss this criterion in more
detail in the coming sections.

\section{Results}

\begin{figure}[t] 
\epsfig{figure=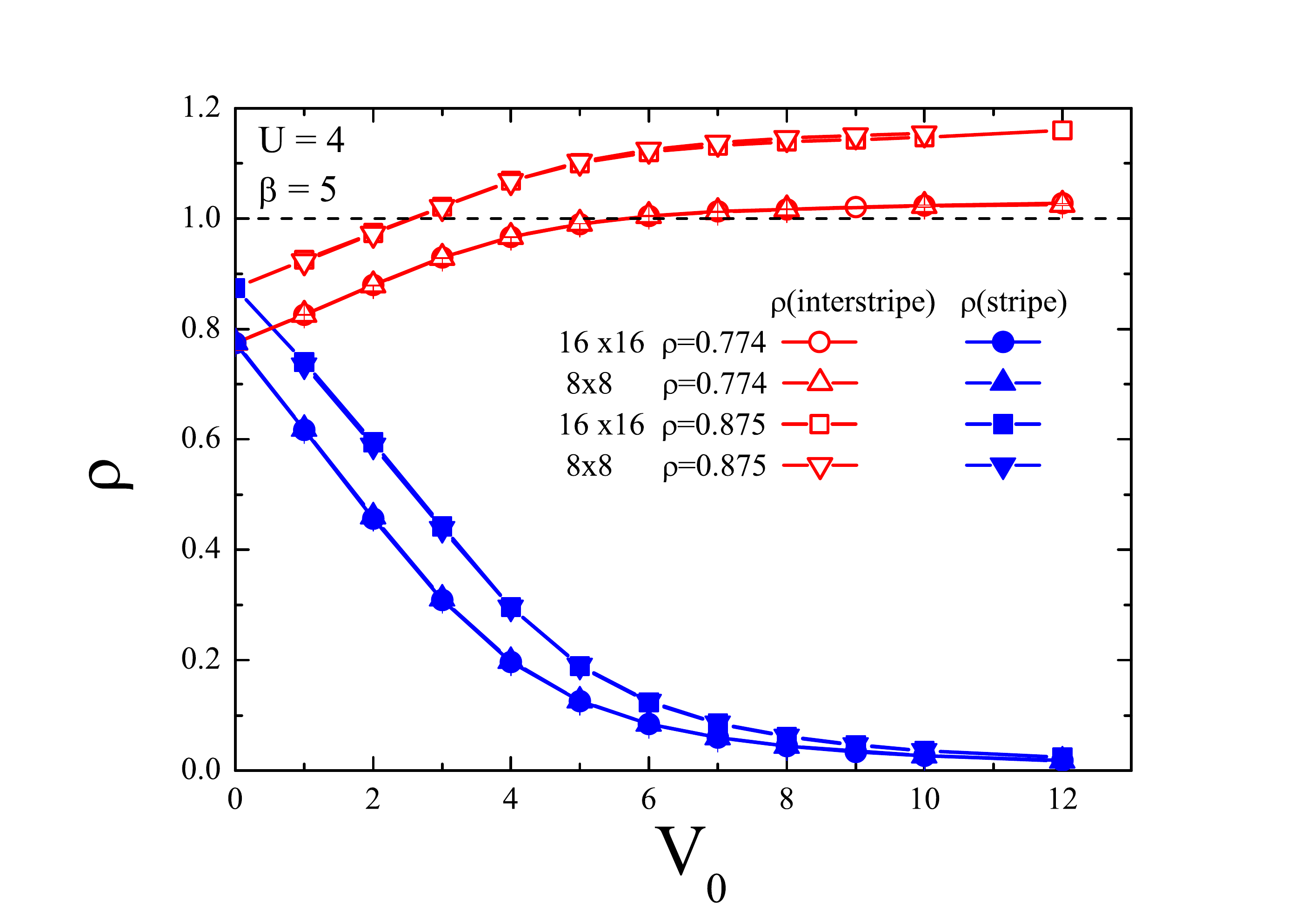,
width=8.0cm,clip}
\vspace{-0.5cm}
\caption{
(Color online) The density of particles is shown on the striped rows where
$V_0$ acts and the unstriped rows.
The total density of the system is fixed at $\rho=0.774$ (circles for 16x16  and up triangles for 8x8 lattices),
and $\rho=0.875$ (squares for 16x16  and down triangles for 8x8 lattices),
There is only a very small variation of density
on the unstriped rows with different distance
from the stripes, so only the average is shown.
Data for 8x8 and 16x16 lattices are
essentially indistinguishable.
\label{rhoV0insideandoutsidestripes}
}
\end{figure}

When the total density, averaged over the entire lattice,
is fixed, the densities on and in-between
the stripes depend on $V_0$ is shown in
Fig.~\ref{rhoV0insideandoutsidestripes} for $\rho=0.774$ and $\rho=0.875$.  $V_0=0$
corresponds to the homogeneous lattice, and the stripe and interstripe densities
are equal there.  As $V_0$ increases, charge is driven
off the stripes until, ultimately, for $V_0 > 10$,
the stripes are nearly empty.  The fermions flow into the
interstripe regions.  Their density rises, going through
half-filling at $V_0 \approx 6$, and asymptotically increasing to a bit over
unity for large $V_0$ for $\rho=0.774$.  If the average density on the entire lattice
were $\rho=3/4$ then, for ${\cal P}=4$, the interstripe regions
would be precisely half-filled when the stripes are completely empty.
For $\rho=0.875$ the interstripe density crosses half-filling at $V_0 \approx 3$ and
approaches $\rho \simeq 1.2$ as $V_0$ increases.

\begin{figure}[t] 
\vspace{-0.5cm}
\hspace{-1.0cm}
\epsfig{figure=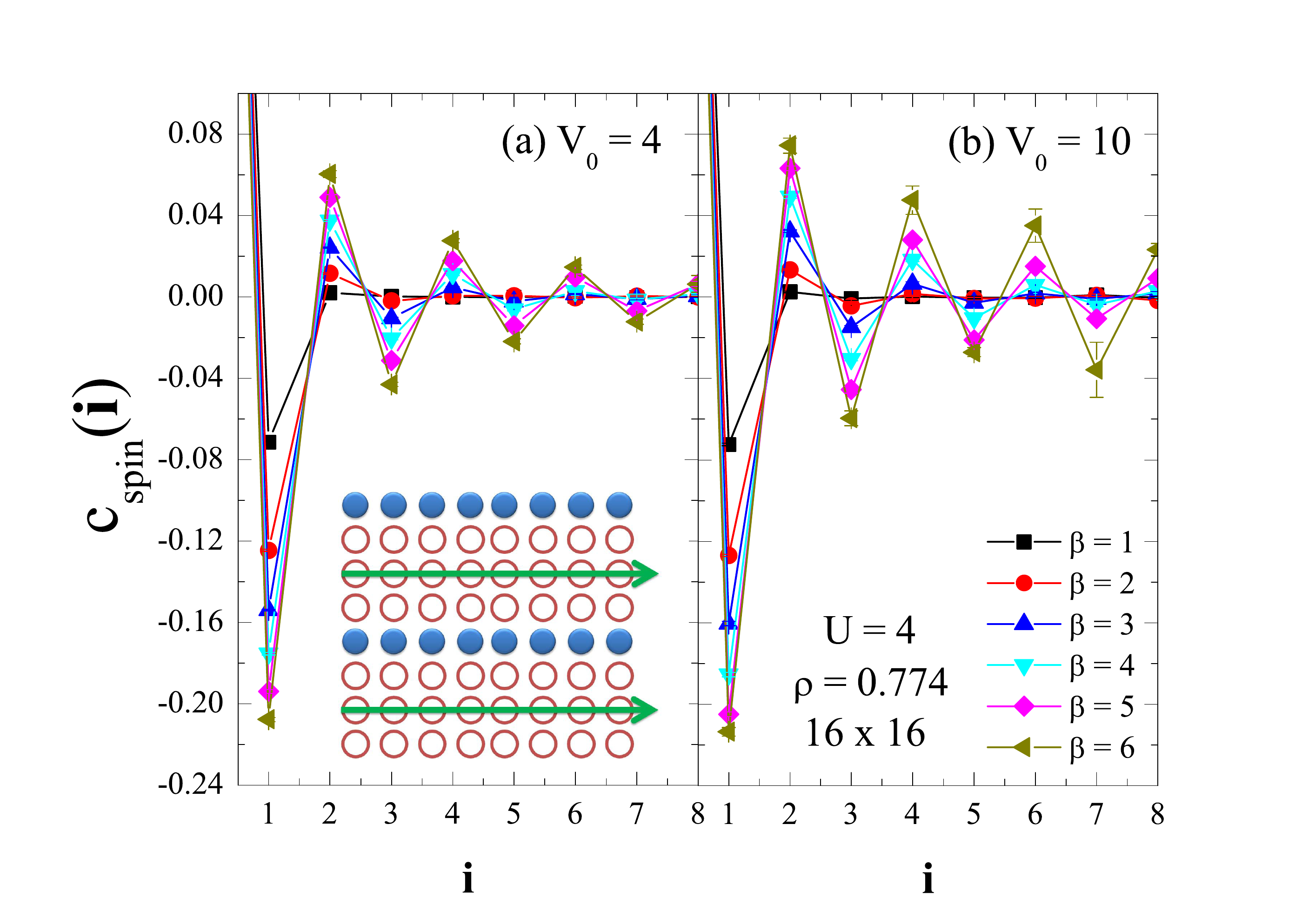,
width=9.5cm,clip}
\vspace{-0.5cm}
\caption{
(Color online) The spin correlation function $c_{\rm spin}({\bf i})$ down the center of
one of the (three site wide) interstripe regions.  Here $U=4$,
and the average density $\rho=0.774$ over the whole lattice, for $V_0=4$ (a) and $V_0=10$ (b).  The
interstripe region is fairly close to half-filling, and so, as the
temperature $T$ is lowered, $c_{\rm spin}({\bf i})$ oscillates over
fairly large distances. The arrows indicate the sites along which the spin-spin correlations are calculated.
\label{spinspinN16U4}
}
\end{figure}

In Fig.~\ref{spinspinN16U4}
the spin-spin correlation $c({\bf i})$ is given down the center of the
interstripe region (i.e.~parallel to the stripes, see arrows in the inset), at $\rho=0.774$
for $V_0=4$ (a) and $V_0=10$ (b).
As seen in
Fig.~\ref{rhoV0insideandoutsidestripes}
these values correspond to densities slightly below and slightly
above half-filling respectively.  Despite the doping, there
is a fairly robust antiferromagnetic pattern as $T$ is lowered.

\begin{figure}[t] 
\vspace{-0.5cm}
\epsfig{figure=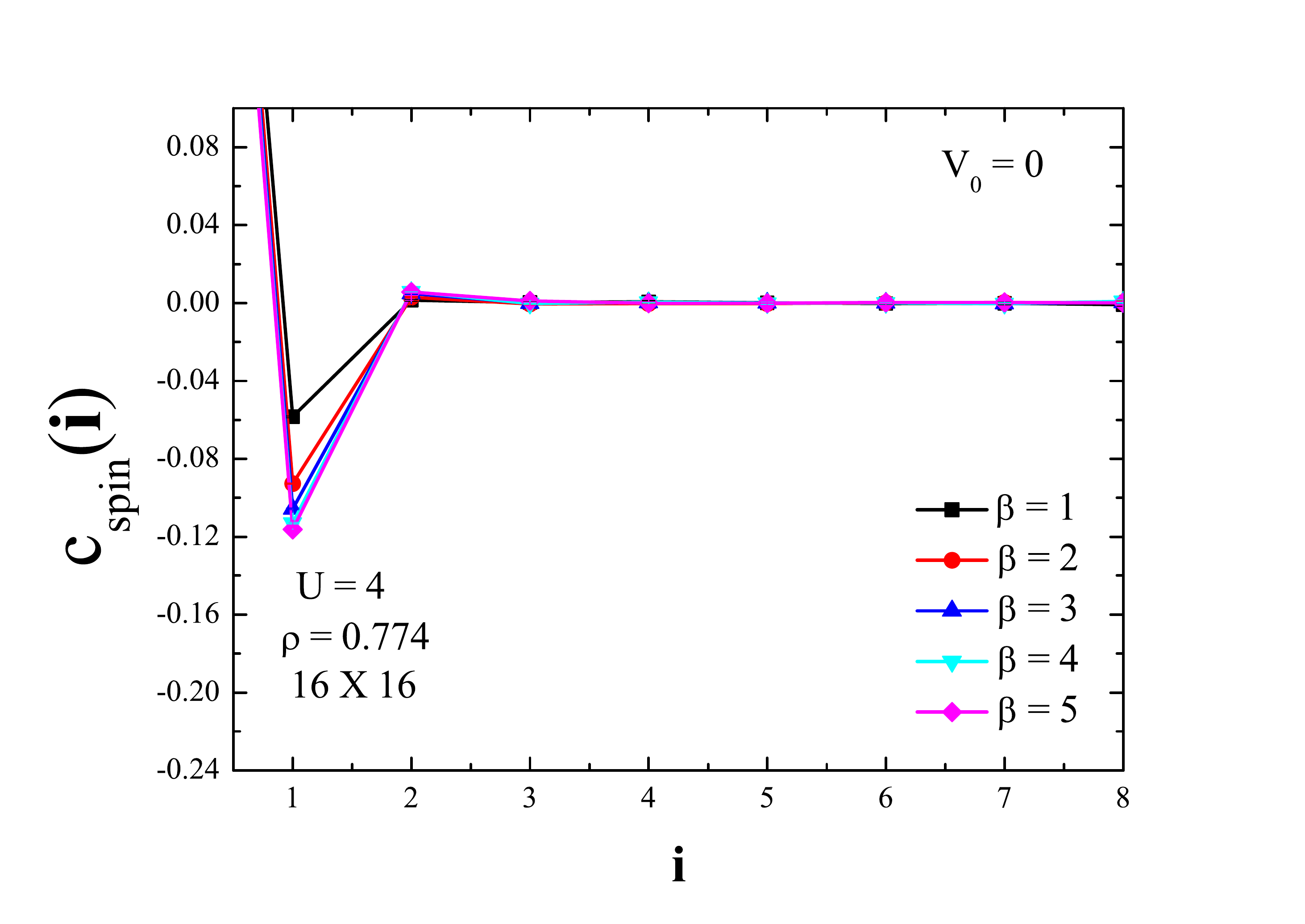,
width=9.0cm,clip}
\vspace{-0.8cm}
\caption{
(Color online) Same as Fig.~\ref{spinspinN16U4} except for $V_0=0$
so that the lattice is at homogeneous density.
When there are no stripes the spin correlations dies out
very rapidly for this doping of the uniform Hubbard Hamiltonian.
They would be unlikely to be able to supply the ``glue" for
Cooper pairing.
\label{spinspinV0_0}
}
\end{figure}

In contrast, in the absence of stripes, $V_0=0$, the doped holes
are spread uniformly throughout the lattice, and for the same density
as exhibited in
Fig.~\ref{spinspinN16U4}
antiferromagnetic order is very short ranged, as seen in Figure
\ref{spinspinV0_0}.
In the absence of
any sort of charge inhomogeneity it would be very unlikely that
these weak magnetic correlations could provide the `pairing glue'
for high temperature superconductivity.  Thus the results of
Figs.~\ref{spinspinN16U4} and \ref{spinspinV0_0}
suggest that domain formation is a prerequisite for any superconductivity
which is postulated to arise from robust magnetism at this density.

\begin{figure}[t] 
\vspace{-0.5cm}
\epsfig{figure=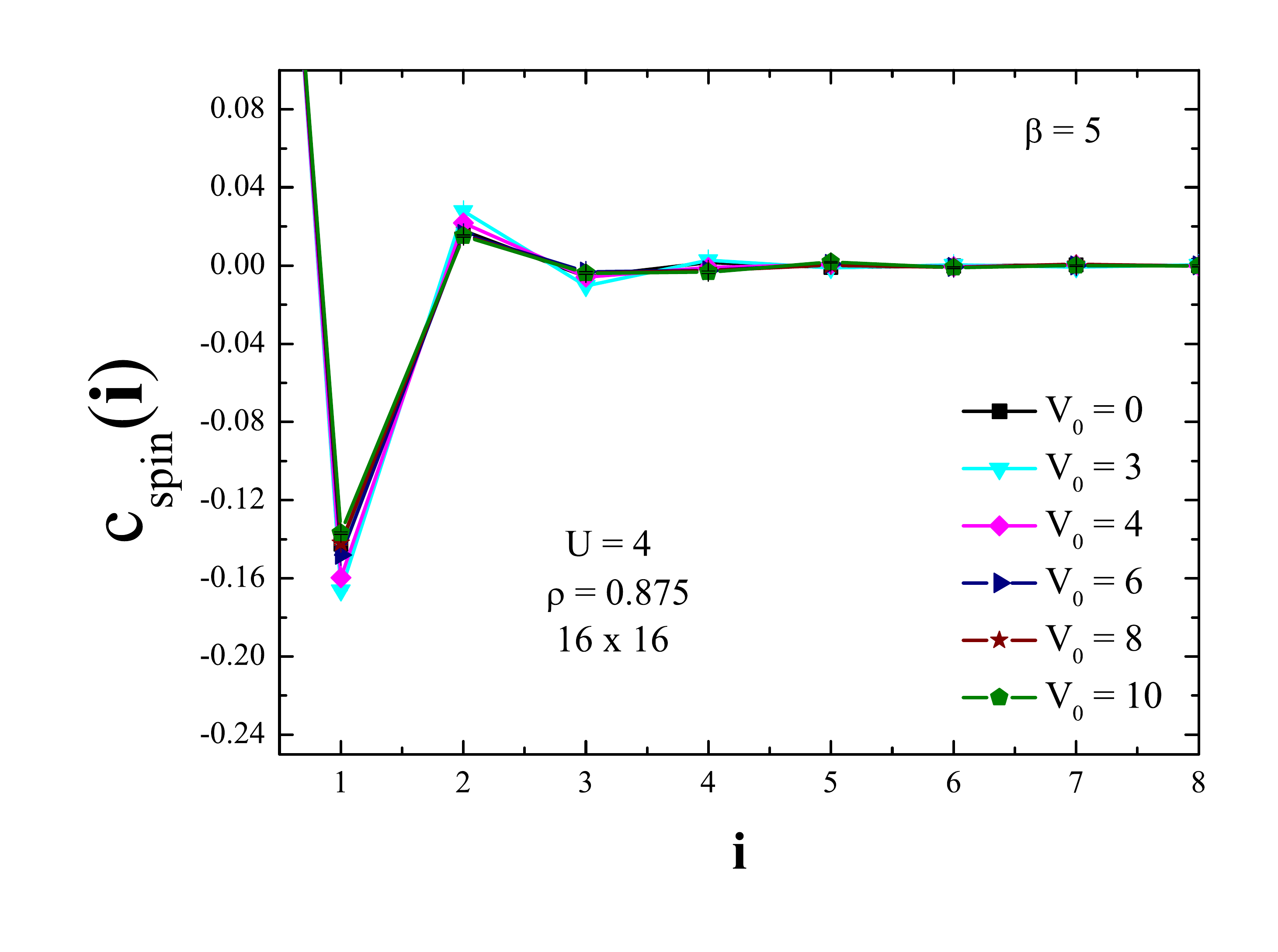,
width=9.0cm,clip}
\vspace{-0.8cm}
\caption{
(Color online) Same as  Fig.~\ref{spinspinN16U4} except for $\rho=0.875$ at fixed $\beta=5$ and different values of $V_0$.
\label{spinspinrho0875}
}
\end{figure}

The presence of stripes alone is not enough to guarantee the presence of antiferromagnetic correlations.  Fig.~\ref{spinspinrho0875} shows the same spin-spin correlations
as in Figs.~\ref{spinspinN16U4}  and \ref{spinspinV0_0}, but  for $\rho=0.875$.
Although spin-spin correlations are slightly higher at $V_0=3$, where the interstripe density is close to one,  antiferromagnetic correlations are very short ranged for all $V_0$.

\begin{figure}[t] 
\vspace{-0.5cm}
\epsfig{figure=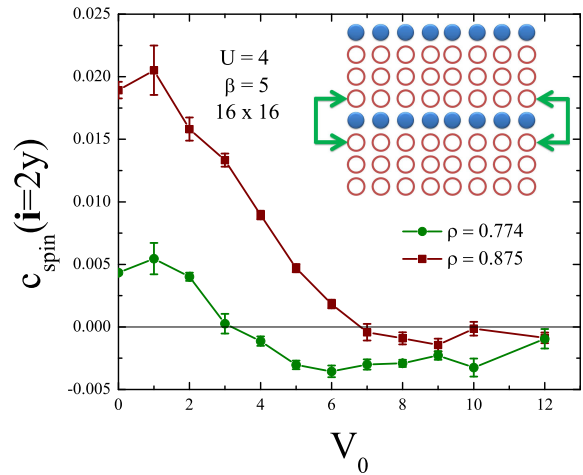,
width=9.0cm,clip}
\vspace{-0.5cm}
\caption{
(Color online) Spin correlations are shown perpendicular to the stripe, and,
specifically for fixed distance ${\bf i}=2\hat y$ which crosses
a stripe and varying $V_0$.
For $V_0$ small (the nearly homogeneous limit),
$c({\bf i}=2 \hat y)$ is positive, as would arise
in an up-down-up-$\cdots$ staggered magnetic pattern.  However,
as stripes are introduced,
$c({\bf i}=2 \hat y)$ flips sign.  The magnetic order
exhibits a `$\pi$ phase shift' and the sublattices of the
(bipartite) square lattice on which the up spin electrons sit are
reversed upon crossing a stripe. Arrows in the inset show pairs of sites traversing the stripe, where the spin-spin correlation functions are calculated.
\label{spincorrelation_Vo}
}
\end{figure}

If the lattice is traversed perpendicular to the stripes, we expect the
spin correlations to be significantly reduced:  the low density on the
stripes does not support a very large moment.
In Fig.~\ref{spincorrelation_Vo}
we show $c({\bf i}=2\hat y)$, corresponding to a pair of sites on a line
parallel to the $\hat y$ axis and traversing a stripe (see inset for a sketch).
The scale of $c({\bf i}=2 \hat y)$ is roughly an order of
magnitude smaller than
$c({\bf i}=2 \hat x)$, as suggested should be the case
by the preceding argument.
Apart from the size of the correlation,
there is another feature of crucial interest.  For $\rho=0.774$ and small $V_0$,
i.e.~close to the homogeneous limit,
the spin correlation two sites away $c({\bf i}=2\hat y)$ is positive,
as expected for an antiferromagnet.
However for $V_0>3$, the sign flips and
$c({\bf i}=2\hat y)$ becomes negative.  This effect is strongly reduced
for $\rho=0.875$, where the negative values of $c({\bf i}=2\hat y)$ are smaller in magnitude
 than those for $\rho=0.774$ and only occur  for $V_0 \gtrsim  7$.

This `$\pi$-phase shift' is
a prominent
experimental feature of stripe physics in the
cuprates \cite{tranquada95,tranquada98}.
These results show that this shift in the sublattice order
across a stripe is also a characteristic of the doped
2D fermion Hubbard model, at least in the case considered here
in which the stripes
are created through an externally imposed potential.

\begin{figure}[t] 
\vspace{-0.5cm}
\epsfig{figure=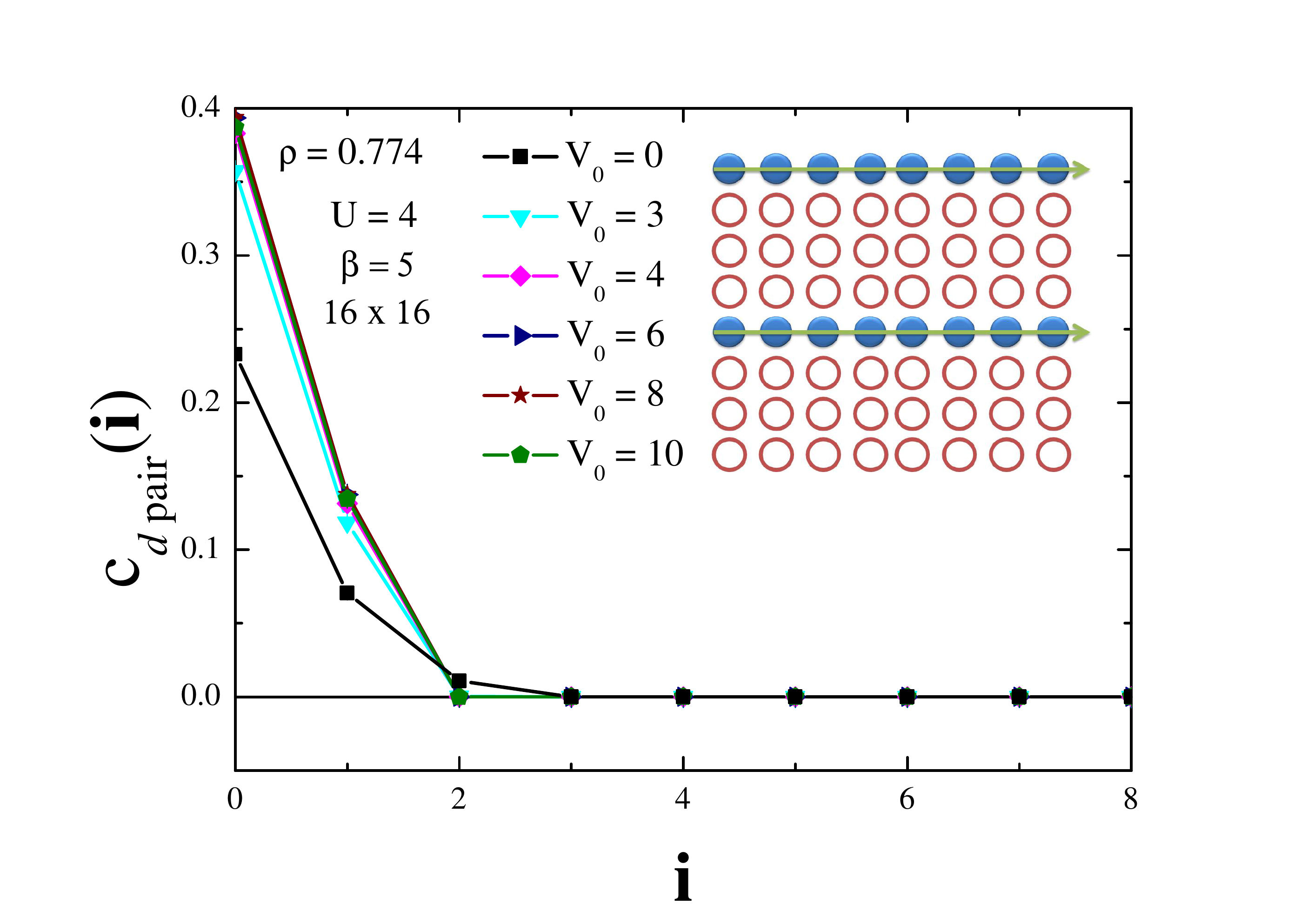,
width=9.0cm,clip}
\vspace{-0.5cm}
\caption{
(Color online) D-wave pair-pair correlation functions along the stripes for $\beta=5$, $\rho=0.774$ and different $V_0$.
Pairing correlations are shown to be short ranged.
\label{dwavealongi}
}
\end{figure}

\begin{figure}[t] 
\vspace{-0.5cm}
\epsfig{figure=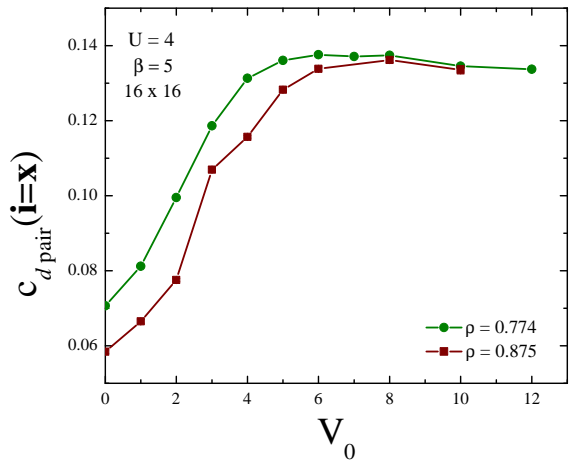,
width=9.0cm,clip}
\vspace{-0.5cm}
\caption{
(Color online) D-wave pair correlations on neighboring sites, along the stripe, as a function of $V_0$. It is clear that,
as $V_0$ is increased the d-wave pairing is enhanced along the stripes.
\label{dwavealong}
}
\end{figure}


We turn now to the pairing correlations.
Figure \ref{dwavealongi} shows the d-wave pairing correlation function along a stripe.
As we will see below, the lowest
temperatures achieved are well above the superconducting transition and therefore the superconducting correlations
are short-ranged.  Fig.~\ref{dwavealong} shows the
pairing correlations between two neighboring sites within a stripe as a function of $V_0$.  It is clear that the presence of stripes
enhances pairing, and it is interesting to note that for both $\rho=0.774$ and $\rho=0.875$, $c_{d \,{\rm pair}}({\bf i})$ stabilizes for $V_0$
close to the value for which the `$\pi$ phase shift' takes place, namely $V_0 \sim 4.0$ and $V_0 \gtrsim 7$, respectively.

\begin{figure}[t] 
\vspace{-0.5cm}
\epsfig{figure=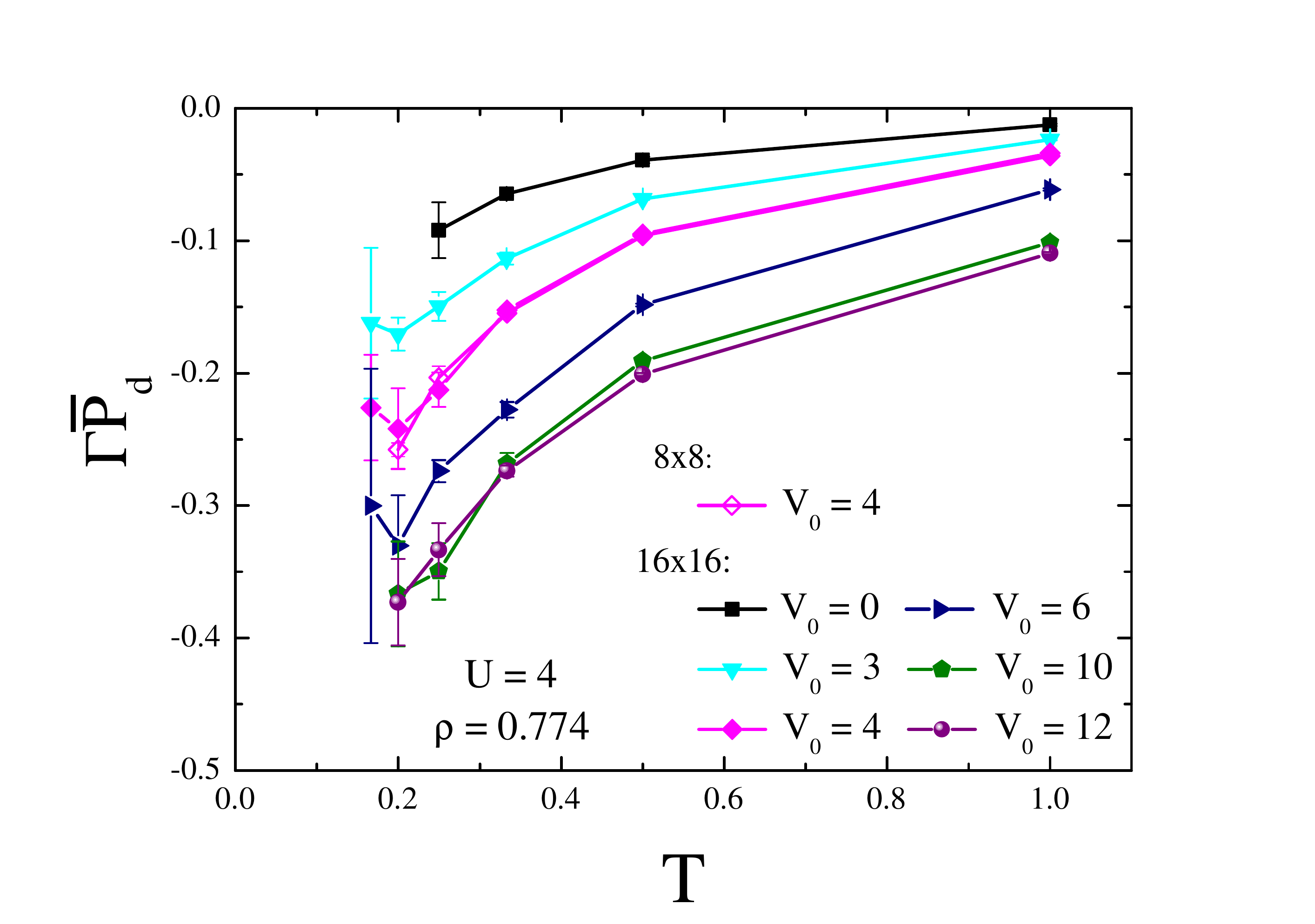,
width=9.0cm,clip}
\vspace{-0.5cm}
\caption{
(Color online) The $d$-wave pairing vertex is shown as a function of temperature
for $U=4$, $\rho=0.774$ and different values of the externally
imposed stripe potential $V_0$.  For the homogeneous system,
$V_0=0$, and small $V_0$ generally, $\Gamma \overline P_d \lesssim
0.10$.
For larger $V_0$, $\Gamma \overline P_d$ exceeds 0.3 in magnitude.
A superconducting instability is signaled by
$\Gamma \overline P_d \rightarrow -1$. Data for 8x8 lattice ($V_0=4$, open symbols)
show finite size effects are negligible.
\label{N16_gammaPdnv}
}
\end{figure}

Figure \ref{N16_gammaPdnv}
shows the key result of this paper.
As charge domains are introduced into the square lattice
Hubbard model, the $d$-wave pairing vertex becomes considerably
more attractive.  Indeed, not only is its magnitude increased by a
factor of three relative to the homogeneous system, the temperature
evolution becomes increasingly steep.  As in all existing DQMC
studies of the repulsive Hubbard Hamiltonian, the sign problem prevents
accessing  low enough temperatures to establish
a critical $T_c$ where $\Gamma \overline{P}_d = -1$,
if such a superconducting transition does indeed occur in this model.
Nevertheless, the results
Figure \ref{N16_gammaPdnv}
are suggestive that charge domains considerably enhance the $d$-wave pairing.

\begin{figure}[t] 
\vspace{-0.5cm}
\epsfig{figure=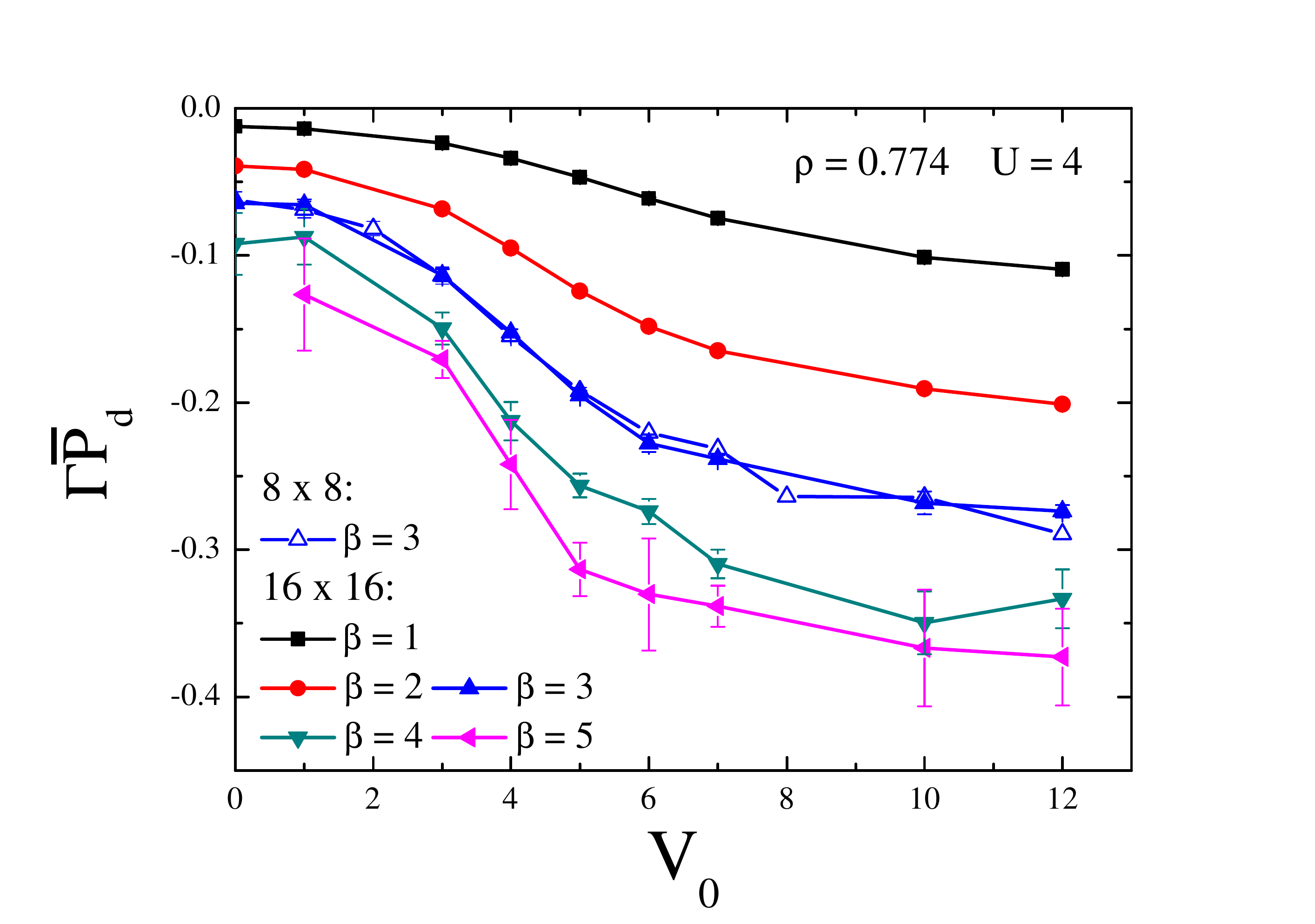,
width=9.0cm,clip}
\vspace{-0.5cm}
\caption{
(Color online) The data of Fig.~\ref{N16_gammaPdnv} are replotted to
show the $d$-wave pairing vertex as a function of $V_0$ for fixed
inverse temperature $\beta$.  $\Gamma \overline{P}_d$ becomes
monotonically more attractive.
Data for 16x16 lattices (closed symbols) and 8x8 lattice (open symbols)
show finite size effects are smaller than statistical fluctuations.
\label{betaall}
}
\end{figure}

We can study this same data as a function of $V_0$ at constant
temperature.  Figure \ref{betaall} shows results for our canonical
parameters, 16x16 lattices, $U=4$ and $\rho=0.774$.
We find that the pairing vertex becomes more and more robust
with increasing $V_0$.  This is not completely intuitive.  One might
expect that a maximum pairing would occur for $V_0 \approx 6$ where,
according to Fig.~\ref{rhoV0insideandoutsidestripes},
the inter-stripe regions are most close to half-filling and
hence antiferromagnetic correlations are most strong.
Alternately, as $V_0$ becomes very large the particle density
within the stripes drops to zero.  If the physical picture is
that of pairing of mobile carriers in the doped region driven
by spin correlations in the half-filled domains, at large $V_0$ the
density of these carriers becomes small, and one would again expect
$\Gamma \overline{P}_d$
to turn over.

Recent DCA calculations\cite{maier10} indeed reveal an initial
enhancement of pairing with the introduction of stripes, followed by a
fall-off as the above physical arguments suggest.  However, this
non-monotonicity is observed only at very low temperatures $T \sim 0.05$
quite close to $T_c$.  For higher $T$, closer to the range studied here,
there is no sign of the $d$-wave eigenvalue coming down at large $V_0$.
Although the lower temperature scales is the most likely explanation for
the DCA non-monotonicity, it is also possible that, in the work
presented here, the decline of pairing correlations in the doped stripes
is compensated by an increase in the interstripe domains, which are
shifted away from half-filling at large $V_0$.  It seems clear that for
large $V_0$ and density $\rho=0.750$ where the interstripe regions are
precisely half filled, the pairing signal would be forced to be small,
and hence that a turnover such as is seen in DCA calculations
should occur.

\begin{figure}[t] 
\vspace{-0.5cm}
\epsfig{figure=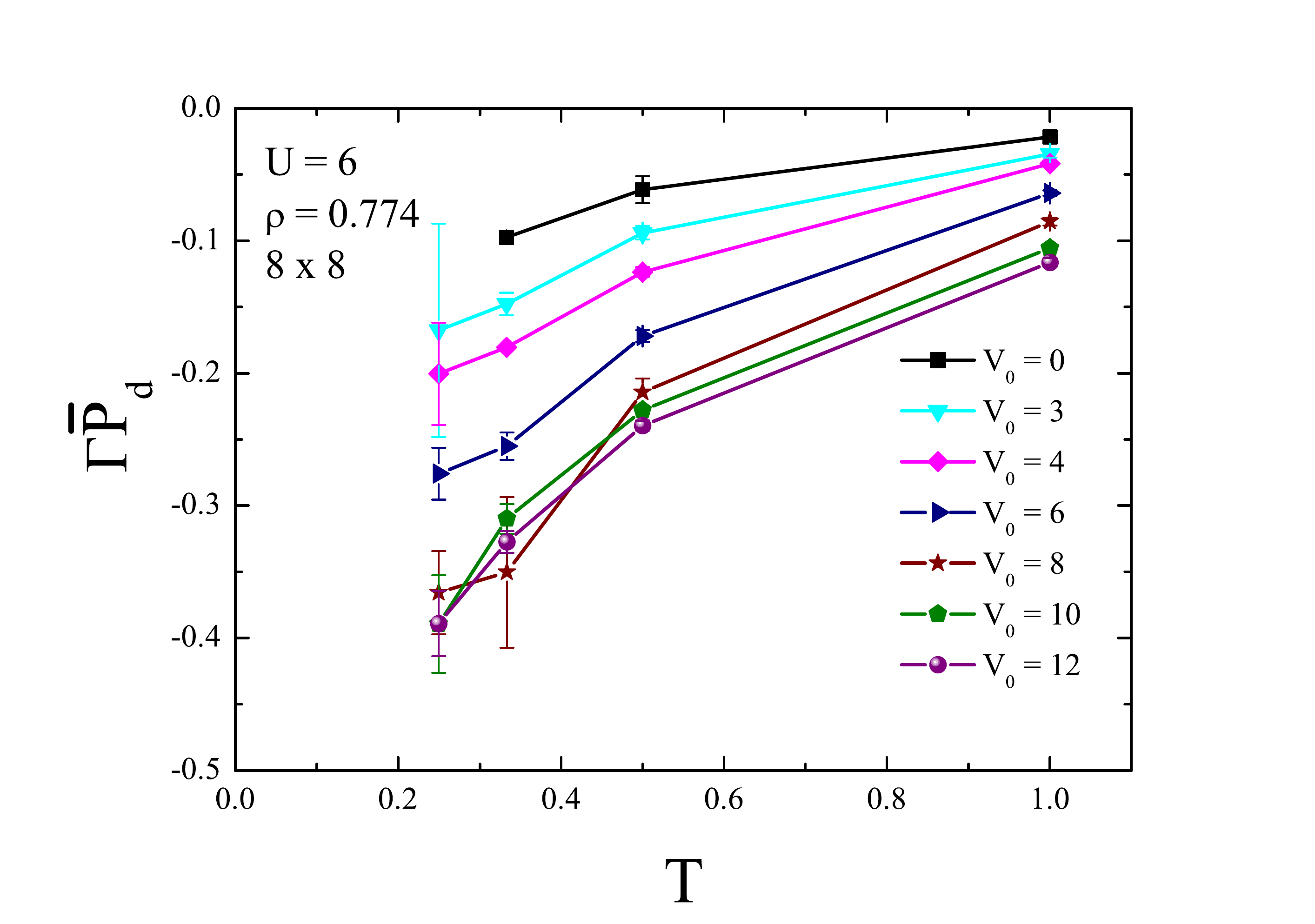,
width=9.0cm,clip}
\vspace{-0.5cm}
\caption{
(Color online)
The pairing vertex is shown as a function of temperature
for the same parameters as Fig.~\ref{N16_gammaPdnv},
except  for a larger on-site repulsion $U=6$ rather than
$U=4$, and 8x8 lattices.  The generic behavior is similar:  charge inhomogeneities
make the $d$-wave vertex more attractive.
\label{gammapdnvU6}
}
\end{figure}

The physics of the half-filled homogeneous Hubbard model on a square
lattice is believed not to be highly sensitive to the interaction
strength $U$.  That is, the ground state is an antiferromagnetic
insulator for all $U$, although the precise nature of the phase evolves
from a weak coupling regime described by spin density wave physics to a
strong coupling Mott insulator.
In order to assess whether the enhancement of $d$ wave pairing
due to striped formation is similarly generic to different $U$
or specific to $U=4$, we show data for
$\Gamma \overline P_d$ as a function of temperature $T$
in Fig.~\ref{gammapdnvU6}
for $U=6$.  The same basic evolution is observed as in
Fig.~\ref{N16_gammaPdnv}, with the vertex being enhanced by $V_0$
both in magnitude and also in the steepness of its
evolution as $T$ is lowered.
A comparison of the data ranges of
Figs.~\ref{N16_gammaPdnv} and \ref{gammapdnvU6}
also reveals some of the limitations of DQMC.  As $U$ gets larger the
sign problem grows, as do the fluctuations (error bars).  For $U=6$
the lowest accessible temperature is $T \sim 0.3$ ($\beta=3$), compared to
$T \sim 0.2$ ($\beta=5$) at $U=4$.
Smaller interaction strengths, e.g.~$U \sim 2$, can readily
be simulated, but tend to harbor large finite size effects which
are the remnants of high degeneracies in the noninteracting
energy levels on a square lattice.

\begin{figure}[t] 
\vspace{-0.5cm}
\epsfig{figure=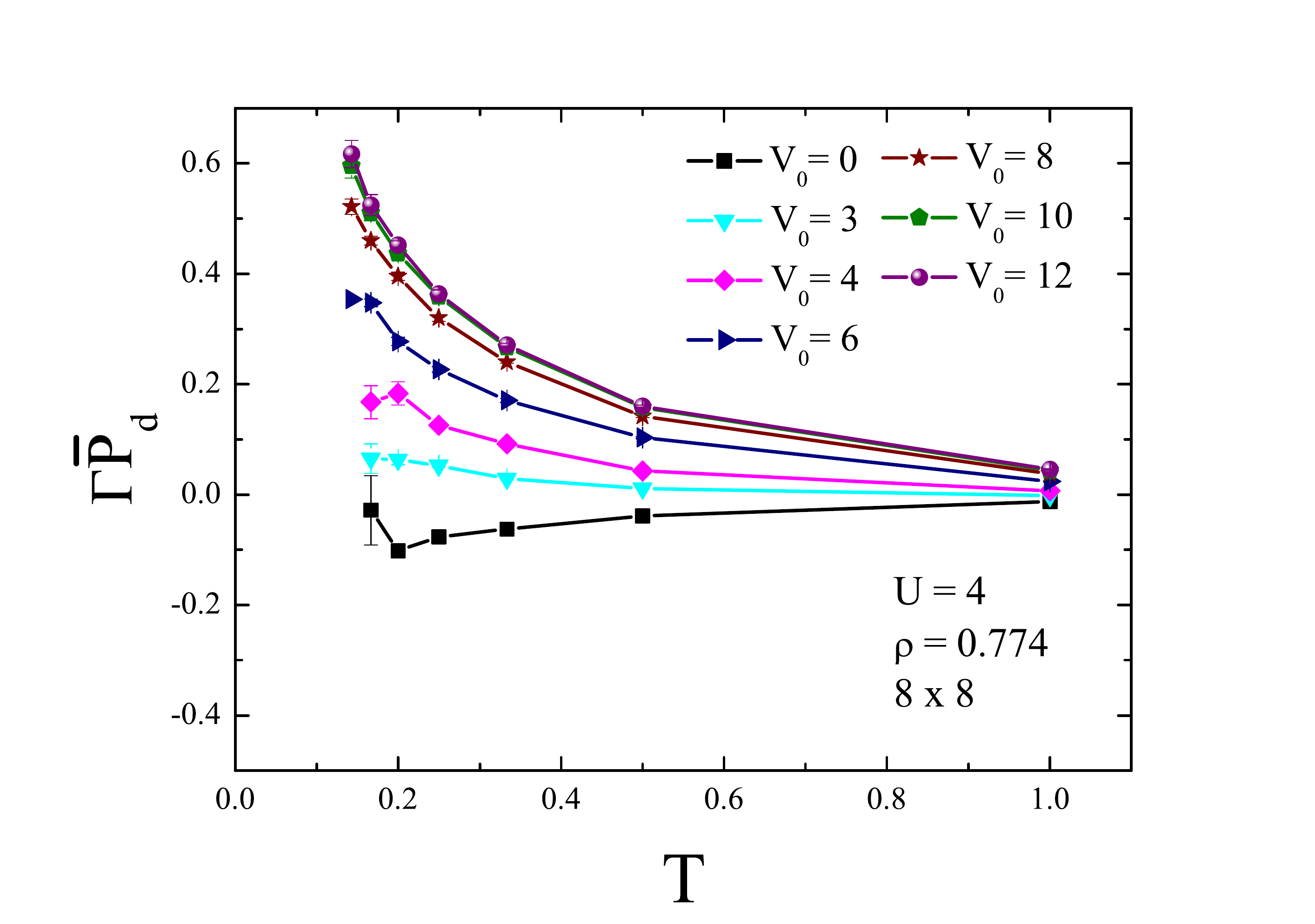,
width=9.0cm,clip}
\vspace{-0.5cm}
\caption{
(Color online)
Pairing vertex for
the ${\cal P}=2$ case, where the stripes are separated by a single
interstripe row.
Unlike the ${\cal P}=4$ case where a three-site-wide interstripe region
separates the stripes, increasing charge inhomogeneity
is detrimental to pairing.  In fact, the $d$-wave vertex
becomes repulsive for $V_0>2$.
\label{gammapdnv_1-1}
}
\end{figure}

We next explore a different lattice periodicity, ${\cal P}=2$.
Unlike the case of different interaction strengths, where the
qualitative physics remains unchanged,
${\cal P}=2$ behaves in a very different manner from
${\cal P}=4$,
as seen in Fig.~\ref{gammapdnv_1-1}.
In this case, increasing $V_0$ reduces the magnitude of
$\Gamma  \overline P_d$,
and by the time $V_0=3$ the vertex has even changed sign
and become repulsive.  It should be noted that because
for ${\cal P}=2$ there
are fewer stripes which do not feel the $V_0$ potential
to absorb the fermions as $V_0$ increases, their density
is increased above $\rho=1$, where AF correlations are most
evident, far more easily than in  the ${\cal P}=4$ case shown in
Fig.~\ref{rhoV0insideandoutsidestripes}.
This is the probable cause for the decrease in $d$ wave pairing
correlations.

\begin{figure}[t] 
\vspace{-0.5cm}
\epsfig{figure=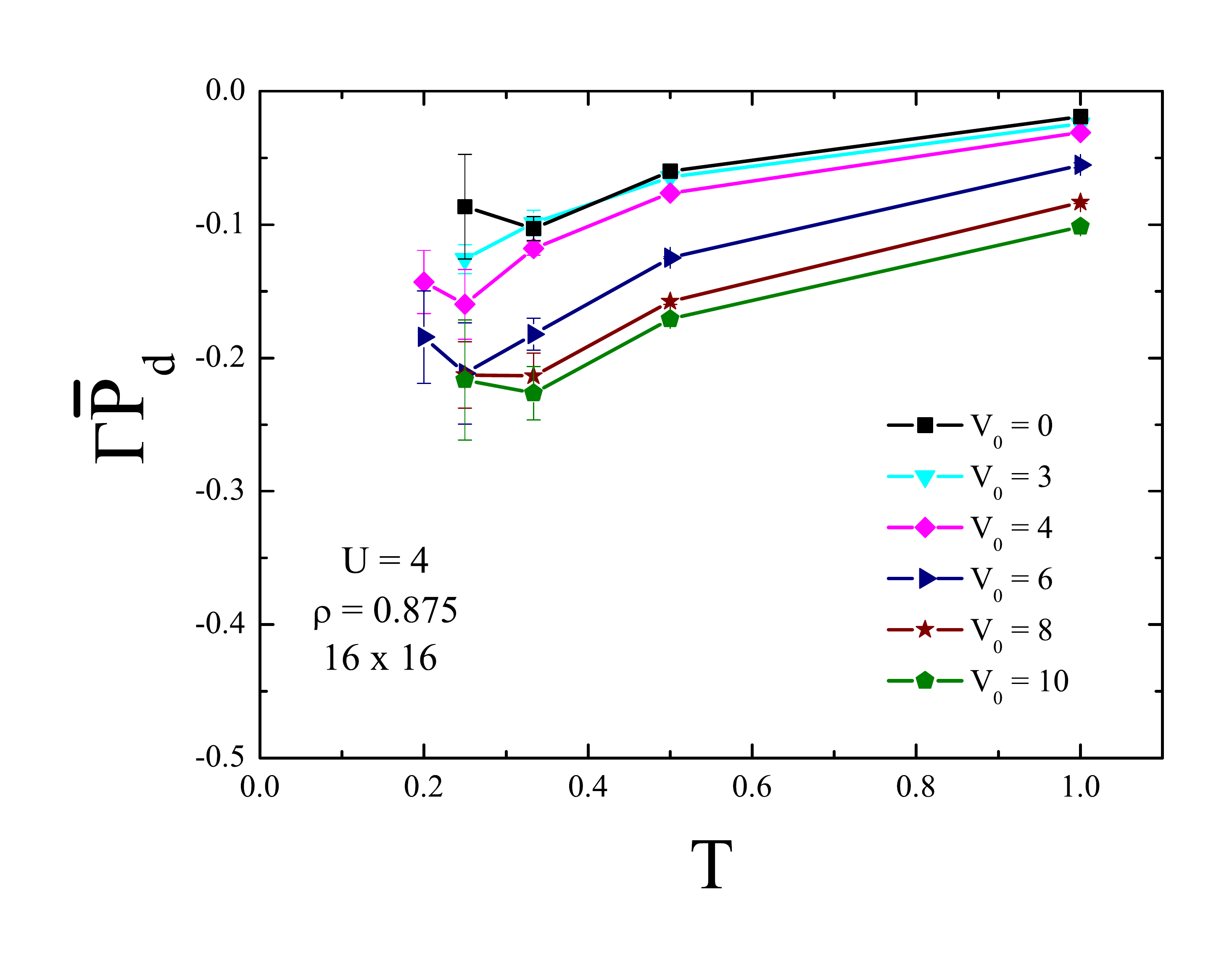,
width=9.0cm,clip}
\vspace{-0.5cm}
\caption{
(Color online)
For a total density corresponding to close to
the `optimal' doping of the cuprate superconductors,
$\rho=0.875$, the pairing vertex, while remaining attractive,
shows somewhat less enhancement as $V_0$ is turned on.
\label{gammapdnv_0875}
}
\end{figure}

Our final check of the robustness of the enhancement of pairing
by stripes is to explore a different density $\rho=0.875$,
which is optimal for pairing in the absence of stripes.
Fig.~\ref{gammapdnv_0875} indicates that the temperature evolution
for non-zero $V_0$ is essentially unchanged from the homogeneous
case.  Thus, even though at $V_0=0$ the product
$\Gamma \overline P_d$ is larger for $\rho=0.875$ than for
$\rho=0.774$, this is no longer the case for nonzero $V_0$.
Indeed, stripes can ultimately make the pairing vertex substantially larger
for the lower density. The strong antiferromagnetic correlations
induced by the stripes at ${\cal P} =4$ provide the ``glue" for pairing at $\rho=0.774$, whereas
at $\rho=0.875$ this ``glue" is not present.

\begin{figure}[t] 
\centerline{\epsfig{figure=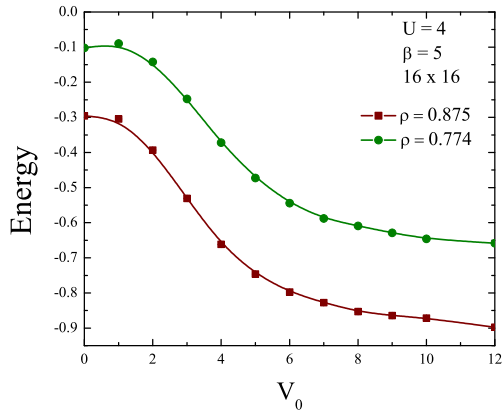,
width=8cm,angle=0,clip}}
\caption{
(Color online)
The total energy  as a function of $V_0$ for 16x16 lattices with $U=4$, $\beta=5$, ${\cal P}=4$, and
$\rho=0.875$ (squares) and $\rho=0.774$ (circles).
\label{energyv0}
}
\end{figure}

\begin{figure}[t] 
\centerline{\epsfig{figure=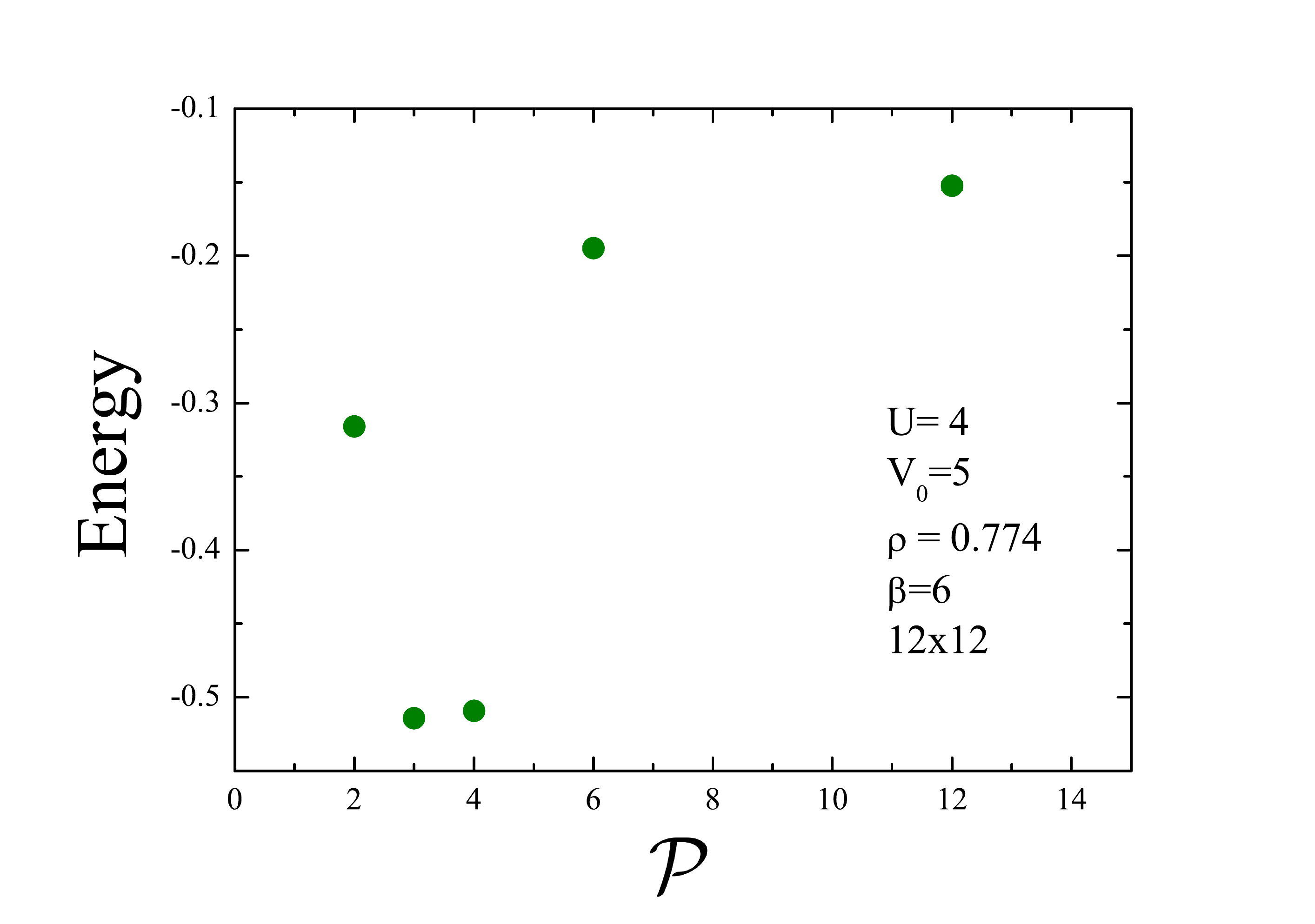,
width=8cm,angle=0,clip}}
\caption{
(Color online)
The total energy for different distances between stripes.
The minimum for ${\cal P}=3,4$ is associated with the fact that
the density of the $V_0=0$ rows
is close to half-filling  for total density
$\rho = 0.774$.
\label{totalenergystripes}
}
\end{figure}

We have focussed thus far on the density, spin, and pairing
correlations. Fig.~\ref{energyv0} examines the total energy as a function of $V_0$ on 16x16 lattices.
Although we have imposed $V_0$ in our Hamiltonian, this computation of
the energy provides a crude measure of the tendency for spontaneous
stripe formation.  The monotonically decreasing  behavior of the energy with $V_0$, suggests that
a maximization of charge imbalance is favored.
When we study the effect of the periodicity of the stripes,
as shown in  Fig.~\ref{totalenergystripes} on a 12x12 lattice,
we see that ${\cal P}=3,4$ have much lower energies,
suggesting their formation might be favored.  These periodicities have
densities on the lines which are not subject to the additional potential
$V_0$ relatively close to half-filling, and hence they have the largest
antiferromagnetic correlations.

\section{Conclusions}

The external imposition of stripes via the introduction of a linear
pattern of reduced chemical potential has been shown to result
in a significant enhancement of the $d$-wave pairing vertex of the
two dimensional Hubbard Hamiltonian.
When the overall density and periodicity of the stripes are such that the density in the interstripe region is close to one,
antiferromagnetic correlations are also
made larger, and exhibit a $\pi$-phase shift across the stripes.
Both the $\pi$-phase shift and the growth in the superconducting
response occur only when the charge inhomogeneity is sufficiently
large, specifically when the additional inhomogeneous site
potential $V_0$ exceeds roughly three times the hopping.

This enhancement of superconductivity has previously been observed
in the closely related dynamical cluster approximation treatment
of the two-dimensional Hubbard Hamiltonian, again in the case
when a site potential $V_0$ was introduced externally.
In this situation, the modulation was chosen to be broader
than the purely one dimensional pattern explored here.
The observation of an optimal stripe potential in the DCA
calculations might be associated with this difference.
It is also interesting to note the possible differences between
pinned and fluctuating stripes.
There are suggestions\cite{tranquada98,vierti94,zaanen96,eskes96}
that the motion of charge/spin domain walls is important
to the enhancement of superconductivity, whereas frozen stripes,
such as created by Nd doping, is inimical to pairing.
The studies in the present paper, as well as the earlier DCA work of
Maier {\it etal} \cite{maier10} indicate that the situation may not
be quite so straightforward and that, in fact, the stripes produced
by an external potential might also be able to enhance
superconductivity.

We have motivated our form for the externally imposed stripe
potential at ${\cal P}=4$ as producing the charge/spin patterns
suggested by neutron scattering \cite{tranquada95}
and DMRG calculations \cite{white98}.
Our pattern contains equal Fourier components for all
wave vectors $Q_n=2\pi n/{\cal P}$.  In the DCA work \cite{maier10}
the effect of different Fourier components $Q=\pi/2$ and $Q=\pi/4$
was explicitly isolated, with the former showing little effect on
pairing and the latter driving significant enhancement.

Spontaneous stripe formation in the doped Hubbard model, if it occurs,
takes place at temperatures below those accessible to DQMC simulations,
which are limited by the sign problem\cite{loh90} to temperatures
greater than roughly one-fortieth of the noninteracting bandwidth.  It
would be interesting also to explore the possible enhancement of pairing
at these temperature scales by other types of charge inhomogeneities
such as checkerboard patterns\cite{hoffman02}  and in the presence of nonmagnetic disorder\cite{romer12,andersen10}.


Tao Ying was supported by a fellowship from the China Scholarship Council. Support from CNPq and FAPERJ (TP and RM) is greatly acknowledged. This work is supported by NSF-PIF-1005503 and DOE SSAAP DE-FG52-09NA29464. We thank B. Brummels for useful input and Steven A. Kivelson for invaluable comments on the manuscript.



\begin{thebibliography}{100}
\bibitem{foot1}
Coupling to a specific component of spin does not involve any
approximation-  the expectation values of all observables
retain the rotational invariance of the Hamiltonian.
However, the error bars do depend on spin component.
\end{thebibliography}
\end{document}